\documentclass[aps,prb,twocolumn,showpacs,groupedaddress]{revtex4}

\usepackage{graphicx}
\usepackage{latexsym}
\usepackage{amssymb}

\begin{document}

\title{Specific heat of stage-2 MnCl$_{2}$ graphite intercalation compound: co-existence of spin glass phase and incommensurate short-range spin order}

\author{Masatsugu Suzuki }
\email[]{suzuki@binghamton.edu}
%\thanks{}
\affiliation{Department of Physics, State University of New York at Binghamton, Binghamton, New York 13902-6016}

\author{Itsuko S. Suzuki }
\email[]{itsuko@binghamton.edu}
%\thanks{}
\affiliation{Department of Physics, State University of New York at Binghamton, Binghamton, New York 13902-6016}

\author{Tadashi Adachi}
%\email[]{}
%\thanks{}
\affiliation{Department of Applied Physics, Graduate School of Engineering, Tohoku University, Sendai 980-8579, JAPAN}

\author{Yoji Koike}
%\email[]{}
%\thanks{}
\affiliation{Department of Applied Physics, Graduate School of Engineering, Tohoku University, Sendai 980-8579, JAPAN}


\date{\today}

\begin{abstract}
Stage-2 MnCl$_{2}$ GIC magnetically behaves like a quasi 2D $XY$ antiferromagnet on the triangular lattice (in-plane lattice constant of Mn layer is 3.692 $\AA$ and the $c$ axis repeat distance is 12.80 $\AA$). The temperature ($T$) dependence of the zero-field cooled (ZFC) and field-cooled (FC) magnetizations shows a typical spin glass behavior below the spin freezing temperature $T_{SG}$ (= 1.1 K). The AC magnetic susceptibility shows a peak at $T$ = $T_{SG}$ at $H$ = 0. This peak shifts to the low temperature side with increasing $H$, according to the de Almeida-Thouless critical line. We have undertaken an extensive study on the $T$ dependence of specific heat in the absence of an external field $H$ and in the presence of $H$ along the $c$ plane. The magnetic specific heat $C_{mag}$ at $H$ = 0 shows no anomaly at $T_{SG}$, but exhibits double broad peaks around 5 K and 41 K. The anomaly at 41 K is the onset of short range spin correlation. The anisotropy of the DC magnetic susceptibility starts to become appreciable below 50 K. The magnetic specific heat $C_{mag}$ at $H$ = 0 is described by $C_{mag} \propto (1/T^{2})\exp(-\Delta/T)$ with $\Delta = 1.41 \pm  0.03$ K, while $C_{mag}$ at $H$ = 10 kOe is proportional to $T$. The entropy due to the broad peak around 5 K is 1/3 of the total entropy. The residual entropy is 63 \% of the total entropy because of highly frustrated nature of the system. The magnetic neutron scattering indicates that a short range spin order associated with the incommensurate wave vector $\left| {\bf Q}_{incomm}\right|$ ($= 0.522 \AA^{-1}$ at 0.45 K) appears below 5 K and remains unchanged down to 63 mK. The in-plane spin correlation length is only 18 $\AA$ at 0.45 K. The low temperature phase below $T_{SG}$ is a kind of reentrant spin glass phase where the spin glass phase coexists with a short range spin order associated with $\left| {\bf Q}_{incomm}\right|$. 
\end{abstract}

%\keywords{}

\maketitle

%\section{}
%\subsection{}
%\subsubsection{}

% If in two-column mode, this environment will change to single-column
% format so that long equations can be displayed. Use
%\begin{widetext}
%\end{widetext}

\section{\label{intro}Introduction}
In stage-2 MnCl$_{2}$ GIC, the MnCl$_{2}$ layer consists of a three-layer sandwich of Cl-Mn-Cl layers, with the same layered structure as pristine MnCl$_{2}$. The MnCl$_{2}$ layers are separated periodically by two graphite layers in stacks along the $c$ axis. The $c$-axis repeat distance is $12.80 \AA$. The MnCl$_{2}$ layer forms a triangular lattice (lattice constant 3.692 $\AA$) translationally incommensurate with the graphite host but rotationally locked into 30$^\circ$ with respect to it. Stage-2 MnCl$_{2}$ GIC is a suitable prototype for studying the classical two-dimensional (2D) XY-like antiferromagnet on a triangular lattice (AFT). The interplanar interaction between adjacent MnCl$_{2}$ layers, $J^{\prime}$ can be greatly reduced by these intervening graphite layers, while the antiferromagnetic intraplanar exchange interaction between Mn$^{2+}$ spins, $J_{1}$, remains virtually unchanged.\cite{ref01,ref02,ref03,ref04,ref05,ref06,ref07,ref08,ref09,ref10}

The 2D XY-like AFT has received attention from theorists because the spins in it are fully frustrated. The ground state consists of spins on three sublattices forming 120$^\circ$ angles with respect to each other, the $\sqrt{3}\times\sqrt{3} $ spin structure. Because there are two senses to the spin helicity, the ground state has a twofold discrete degeneracy as well as an XY-like continuous degeneracy. Consequently, it is predicted to undergo two phase transitions, one associated with Ising-type symmetry breaking and the other with a Kosterlitz-Thouless (KT) mechanism.\cite{ref11,ref12,ref13} Note that our system has an incommensurate spin structure close to $2\sqrt{3} \times 2\sqrt{3}$ spin structure (see later). In this sence our system is different from an ideal 2D $XY$ antiferromagnet on the triangular lattice. 

The magnetic properties of stage-2 MnCl$_{2}$ GIC have been extensively studied by DC susceptibility,\cite{ref05} the zero-field cooled (ZFC) and field-cooled (FC) magnetization,\cite{ref06} AC magnetic susceptibility,\cite{ref03,ref04} electron spin resonance (ESR),\cite{ref02,ref07} specific heat,\cite{ref06} and magnetic neutron scattering.\cite{ref08,ref09,ref10} The low temperature phase below $T_{SG}$ (= 1.1 K) is a spin glass phase, reflecting the fully frustrated nature of the system. A short range spin order related to the incommensurate spin structure (typically $18 \AA$ in-plane spin correlation length) appears below 5 K, and remains unchanged even below $T_{SG}$. 

In the present paper, we report the result of specific heat measurement of stage-2 MnCl$_{2}$ GIC between 0.45 K and 50 K, with and without an external magnetic field applied along the $c$ plane. We show that the magnetic specific heat shows no anomaly at $T_{SG}$, but exhibits a broad peak around 5 K. It also shows a broad peak around 41 K, which is the onset of short range spin correlation. There is a relatively large residual entropy at $T$ = 0 K reflecting the fully frustrated nature of the system. The results of specific heat measurements will be discussed in light of the experimental results of the ZFC and FC magnetization, AC magnetic susceptibility, magnetic neutron scattering, ESR, and DC magnetic susceptibility. To this end, the previous data are extensively re-analyzed. Finally we will discuss the origin of the short range spin order related to the incommensurate spin structure. This structure is close to the commensurate structure with the periodicity of ($2\sqrt{3} \times 2\sqrt{3}$).

Recently the magnetic properties of 2D antiferromagnet on the triangular lattice, NiGa$_{2}$S$_{4}$ ($S$ = 1 for Ni$^{2+}$) have been extensively studied.\cite{ref14,ref15,ref16} There are several features which is common to stage-2 MnCl$_{2}$ GIC, because of the same universality class. (i) The incommensurate short range spin order with nanoscale correlation is observed in the temperature range between 1.5 K and 15 K. (ii) The spin glass phase appears below $T_{SG}$ (= 10 K). (iii) The magnetic specific heat shows no anomaly at $T_{SG}$. It shows double broad peaks at 13 and 80 K. (iv) The magnetic specific heat is proportional to $T^{2}$ in the low temperature side. (v) The entropy due to the broad peak around 13 K is 1/3 of the total entropy. It is surprising that in spite of difference in the detail of magnetic and structural parameters, the magnetic behavior of stage-2 MnCl$_{2}$ GIC is very similar to that of NiGa$_{2}$S$_{4}$. This may imply that these common features are a key to understanding the physics of 2D $XY$-like or Heisenberg-like antiferromagnet on the triangular lattice. 

\section{\label{exp}EXPERIMENTAL PROCEDURE}
The sample of stage-2 MnCl$_{2}$ GIC was prepared from kish graphite by vapor reaction of the powdered MnCl$_{2}$ in a chlorine-gas atmosphere of 740 Torr. The reaction was continued at $500^\circ$C for one month. The stoichiometry of the sample was determined from the weight-uptake measurement as C$_{12.70}$MnCl$_{2}$. The specific heat was measured by the thermal-relaxation method at low temperatures down to 0.4 K in magnetic fields parallel to the $c$ plane up to 20 kOe on field cooling, using a commercial apparatus (PPMS, Quantum Design). 

\section{\label{result}RESULT}
\subsection{\label{resultA}Determination of nonmagnetic specific heat at $H=0$}

\begin{figure}
\includegraphics[width=8.0cm]{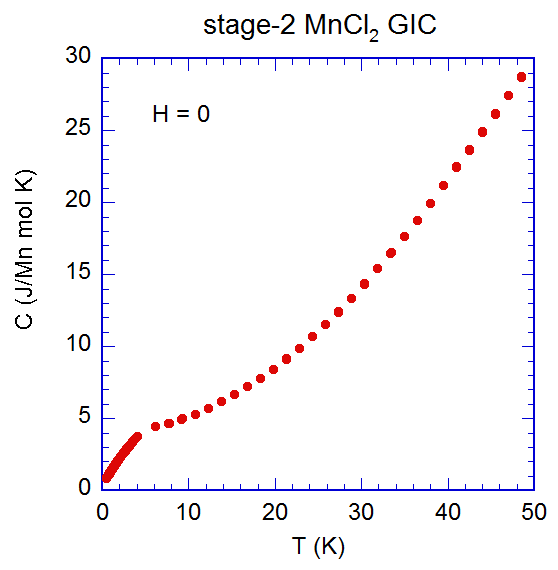}
\caption{\label{fig01}(Color online)Plot of specific heat $C$ in the units of J/(Mn mol K) vs $T$ for stage-2 MnCl$_{2}$ GIC. $H = 0$. Note that two kinds of data which were separately taken under the same condition, are shown.}
\end{figure}

Figure \ref{fig01} shows the $T$ (temperature) dependence of specific heat ($C$) in zero field for a stage-2 MnCl$_{2}$ GIC based on single-crystal kish graphite, where the specific heat is expressed in the unit of J/(Mn mol K). The specific heat increases with increasing $T$ at low temperatures below 5 K, becoming to flatten out between 5 K and 20 K, and it increases with further increasing $T$. There is no anomaly in $C$ vs $T$ around the spin freezing temperature $T_{SG}$ (= 1.1 K), where the ZFC magnetization exhibits a relatively sharp peak. 

\begin{figure}
\includegraphics[width=8.0cm]{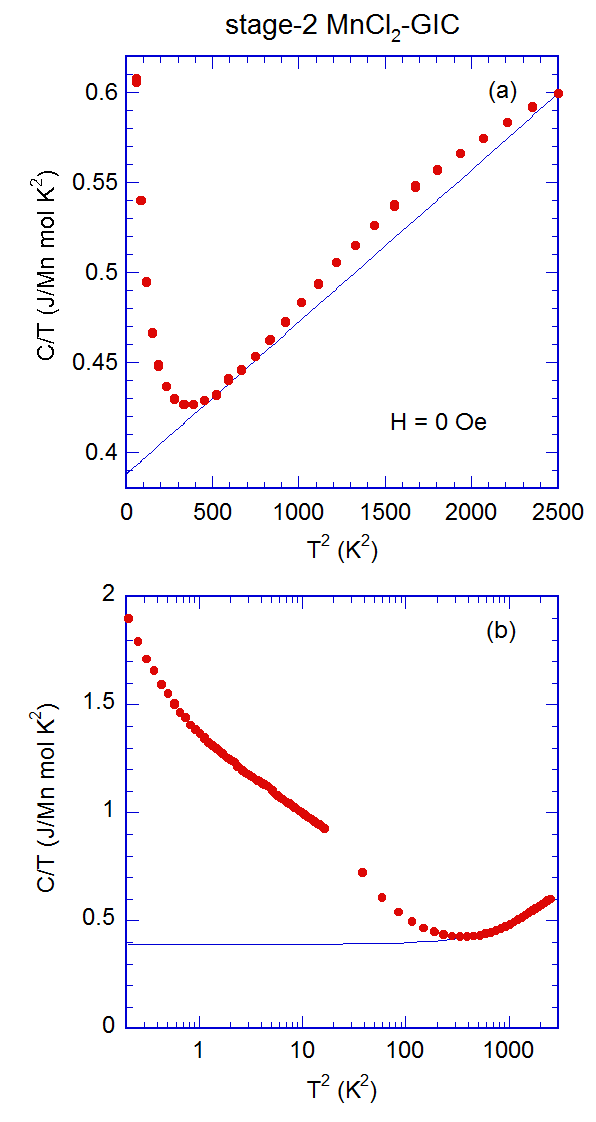}
\caption{\label{fig02}(Color online)(a) $C/T$ vs $T^{2}$ for stage-2 MnCl$_{2}$ GIC. $H = 0$. (b) Overview of the plot of $C/T$ vs $T^{2}$ for stage-2 MnCl$_{2}$ GIC. $H = 0$. The solid line indicates the $C_{nonmag}/T$ vs $T^{2}$, which is obtained from the procedure described in the text.}
\end{figure}

In Figs.~\ref{fig02}(a) and (b), we make a plot of $C/T$ vs $T^{2}$ at $H$ = 0 for stage-2 MnCl$_{2}$ GIC. It is found from Fig.~\ref{fig02} that $C/T$ drastically decreases with increasing $T^{2}$ at low temperatures below 20 K. It shows a minimum around 20 K and in turn it increases with further increasing $T^{2}$. 

It is expected that the specific heat of stage-2 MnCl$_{2}$ GIC at low temperatures, $C$, is described by, 
\begin{equation}
C=C_{mag}+C_{nomag} ,
\label{eq01}
\end{equation}
with
\begin{equation}
C_{nonmag}=C_{e}+C_{g}=\gamma T+\beta T^{3},
\label{eq02}
\end{equation}
where $C_{e}$ (=$\gamma T$) and $C_{g}$ ($=\beta T^{3}$) represent the contributions of the conduction electrons and of the lattice to the specific heat, respectively, and $C_{mag}$ is the magnetic contribution. Here we assume that the Deby temperature of stage-2 MnCl$_{2}$ GIC with stoichiometry C$_{12.70}$MnCl$_{2}$ is on the same order as that of the pristine graphite. Therefore the $T^{5}$ term of the $C_{nonmag}$ may be negligibly small for $T<50$ K. Then $C/T$ is written as
\begin{equation}
\frac{C}{T}=\frac{C_{mag}}{T}+\frac{C_{nomag}}{T}=\frac{C_{mag}}{T}+\gamma +\beta T^{2}  .
\label{eq03}
\end{equation} 
In the temperature range where the magnetic contribution $C_{mag}/T$ is negligibly small, the curve of $C/T$ vs $T^{2}$ is a straight line with the $y$-intercept $\gamma$ and the slope $\beta$. 

In Fig.~\ref{fig02}(a), we note that the data of $C/T$ vs $T^{2}$ for $20 \le T\le 50$ K do not fall on a straight line, but is upward convex for any possible straight line corresponding to the plot of $C_{nonmag}/T$ vs $T^{2}$. This may imply that the magnetic specific heat might exhibit a small broad peak centered around $40 - 50$ K. In fact the existence of the small broad peak has been confirmed from the AC calorimetry measurement by Koga and Suzuki.\cite{ref02} Such a slight deviation of $C/T$ vs $T^{2}$ for $20 \le T\le 50$ K from a straight line makes it difficult for one to determine exactly the $T$ dependence of $C_{nonmg}$. Here we assume that the straight line for the plot $C_{nomag}/T$ vs $T^{2}$ passes two points at $T$ = 22.8 K and 50.5 K. It is expressed by 
\begin{equation}
\frac{C_{nonmag}}{T} =\gamma +\beta T^{2},
\label{eq04}
\end{equation}
with the parameters $\gamma$ and $\beta$ given by $\gamma = 0.38807$ (J/Mn mol K$^{2}$), and $\beta = 8.4397 \times 10^{-5}$ (J/Mn mol K$^{4}$). The advantage of the choice of the straight line for $C_{nonmg}/T$ vs $T^{2}$ is that the plot of $C/T$ vs $T^{2}$ is always above the straight line for $C_{nonmg}/T$ vs $T^{2}$. In other words, $C_{mag}$ is always positive. The disadvantage is that the $T$ dependence of $C_{mag}$ well above 20 K is rather sensitive to the choice of the straight line. Fortunately, the $T$ dependence of $C_{mag}$ well below 20 K, is not so sensitive to the choice of straight line, since the magnitude of $C_{mag}$ is much larger than that of $C_{nonmag}$. 

\subsection{\label{resultB}Magnetic specific heat $C_{mag}$ vs $T$}

\begin{figure}
\includegraphics[width=8.0cm]{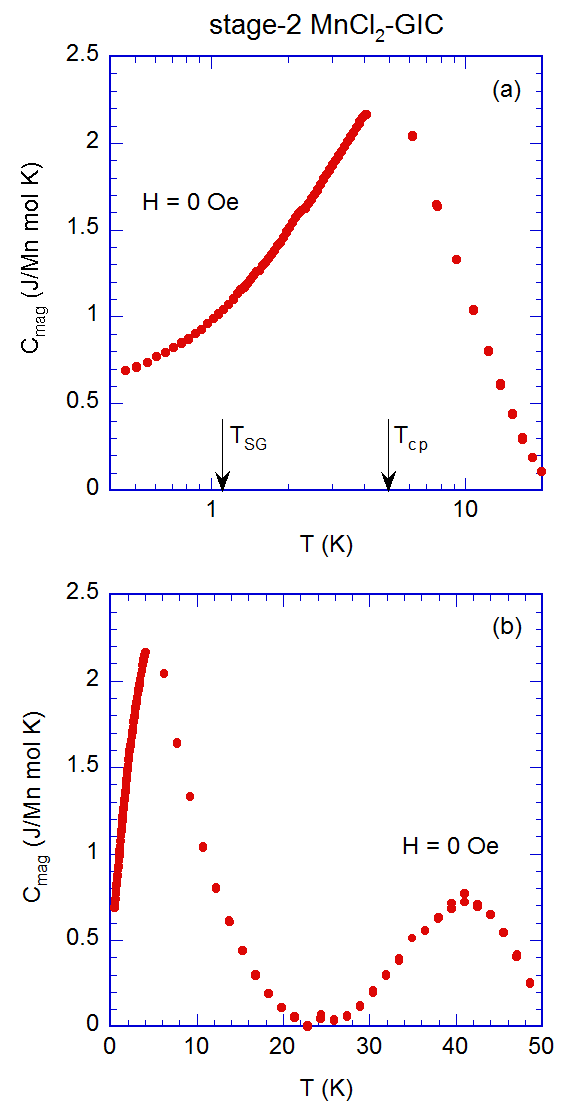}
\caption{\label{fig03}(Color online)Plot of $C_{mag}$ vs $T$ for $T<15$ K. $C_{mag}$ shows a broad peak well above $T_{SG}$ (= 1.1 K). $C_{mag}$ exhibits no anomaly at $T = T_{SG}$, but exhibits a broad peak around $T_{cp}$ ($\simeq 5$ K). (b) Overview of the curve of $C_{mag}$ vs $T$ for $0<T<50$ K. $C_{mag}$ shows double broad peaks around 5 and 41 K. The height of the peak around 41 K may be dependent on the choice of $C_{nonmag}$ as a background. Since $C_{nonmag}$ is very small at low temperatures below $T_{SG}$, the $T$ dependence of $C_{mag}$ is independent of the choice of $C_{nonmag}$.}
\end{figure}
 
In Fig.~\ref{fig03}(a) we show the $T$ dependence of $C_{mag}$ for $H = 0$ at low temnperatures, from the subtraction $C_{nonmag}$ from $C$. Since $C_{nonmag}$ is very small at low temperatures, the uncertainty of $C_{mag}$ is negligibly small. The magnetic specific heat $C_{mag}$ shows a broad peak around 5 K. No anomaly in $C_{mag}$ vs $T$ is observed at the spin freezing temperature $T_{SG}$ (= 1.1 K). Above $T_{cp}$ ($\simeq 5$ K), $C_{mag}$ decreases as $1/T$ with increasing $T$. In Fig.~\ref{fig03}(b) we show the $T$ dependence of $C_{mag}$ between 0.4 and 50 K. The magnetic specific heat exhibits a local minimum around 22 K, and a local maximum around 41 K. The peak height of $C_{mag}$ around 41 K is dependent on the choice of $C_{nonmag}$ as a background, since the contribution of $C_{nonmag}$ to the specific heat $C$ is large. The anomaly at 41 K is closely related to the change of the spin symmetry from isotropic to $XY$ like. Wiesler et al.\cite{ref05} have reported the $T$ dependence of the DC magnetic susceptibility ($\chi_{\perp}$ along the $c$ plane and $\chi_{\parallel}$ along the $c$ axis). Both $\chi_{\perp}$ and $\chi_{\parallel}$ decrease monotonically with increasing $T$ and exhibits no anomaly around 50 K. Although $\chi_{\perp}>\chi_{\parallel}$ at all temperatures, the anisotropy $\Delta\chi=\chi_{\perp}-\chi_{\parallel}$ becomes appreciable only below 50 K. This may indicate the onset of short-range spin correlation in MnCl$_{2}$ layer.

\subsection{\label{resultC}$T$ dependence of magnetic specific heat $C_{mag}$ in the presence of $H$}

\begin{figure}
\includegraphics[width=8.0cm]{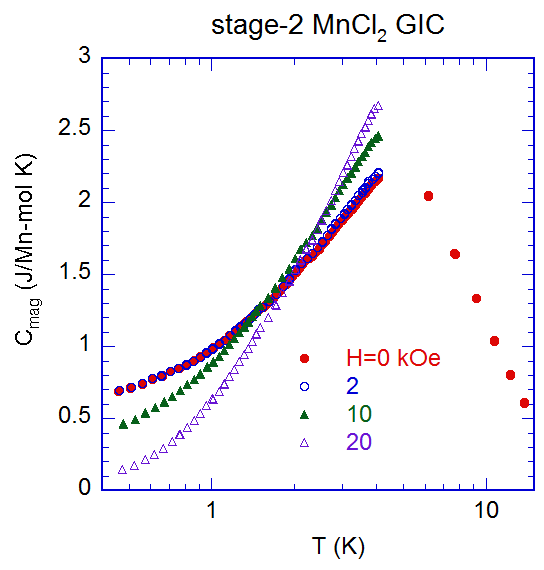}
\caption{\label{fig04}(Color online) Magnetic specific heat $C_{mag}$ vs $T$ at low temperatures for stage-2 MnCl$_{2}$ GIC. An external magnetic field $H$ is applied along the $c$ plane (perpendicular to the $c$ axis). $H$ = 0, 2, 10, and 20 kOe.}
\end{figure}
 
Figure \ref{fig04} shows the $T$ dependence of the magnetic specific heat $C_{mag}$, where an external magnetic field $H$ is applied along the $c$ plane. The external magnetic field $H$ is changed as a parameter; $H$ = 0, 2, 10, and 20 kOe. It is found that $C_{mag}$ is strongly dependent on the magnetic field. As will be discussed in Sec.~\ref{sumC}, when the magnetic field is applied to our system at $T$ below $T_{SG}$, the spin glass phase is destroyed and changes into the paramagnetic phase above the critical field $H_{c}(T)$. This is called the de Almeida-Thouless (AT) transition.\cite{ref17} The critical field $H_{c}(T = 0)$ (defined as $H_{AT}$) is evaluated as $5.87 \pm  0.47$ kOe. This implies that the system is at least in the spin glass phase at low fields ($H = 0$, and 2 kOe).

\subsection{\label{resultD}$T$ dependence of $C_{mag}$ below $T_{SG}$}

\begin{figure}
\includegraphics[width=8.0cm]{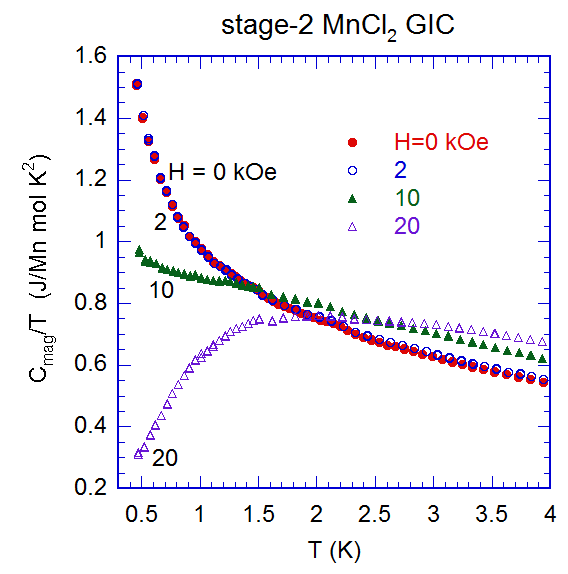}
\caption{\label{fig05}(Color online) Plot of $C_{mag}/T$ as a function of $T$ where $H$ is changed as a parameter. $H$ = 0, 2, 10, and 20 kOe. $H$ is applied along the $c$ plane.}
\end{figure}
 
Figure \ref{fig05} shows the plot of $C_{mag}/T$ vs $T$ for stage-2 MnCl$_{2}$ GIC at low temperatures below 4 K, where $H$ is changed as a parameter. For $H=0$ and 2 kOe, the curve of $C_{mag}/T$ vs $T$ drastically decreases with increasing $T$. For $H=10$ kOe, it gradually decreases with increasing $T$. For $H=20$ kOe, in contrast, the curve of $C_{mag}/T$ vs $T$ starts to increase linearly with increasing $T$ and shows a broad peak around $T=2$ K. 

\begin{figure}
\includegraphics[width=8.0cm]{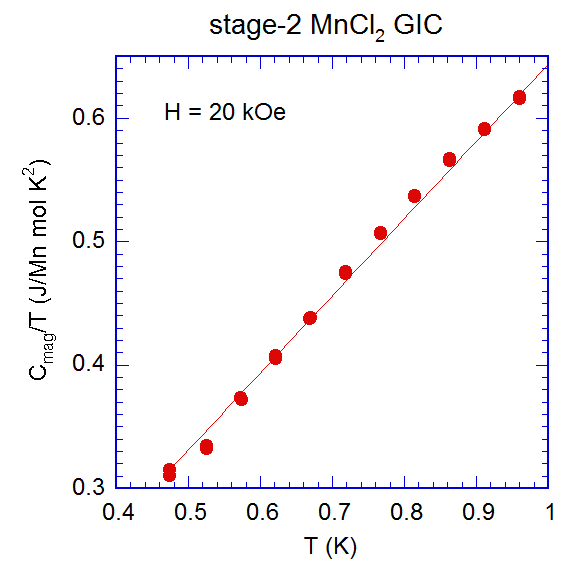}
\caption{\label{fig06}(Color online) Plot of $C_{mag}/T$ vs $T$ for $H = 20$ kOe. $C_{mag}/T$ is proportional to $T$ below $T_{SG}$. The solid line is a least-squares fitting curve.}
\end{figure}

Here we examine the $T$ dependence of $C_{mag}$ below $T_{SG}$ for each $H$. For $H = 20$ kOe, as shown in Fig.~\ref{fig06}, the least-squares fit of the data of $C_{mag}/T$ vs $T$ for 0.45 K $\le T\le T_{SG}$ to
\begin{equation}
C_{mag}/T =\alpha_{0} + \delta_{0} T,
\label{eq05}
\end{equation} 
yields the parameters
\[ 
\alpha_{0} = 0.0186 \pm 0.0088 \text{  J/Mn mol K}^{2}
\] 
and
\[ 
\delta_{0} = 0.6253 \pm 0.0116 \text{  J/Mn mol K}^{3} .
\] 
This implies that the magnetic specific heat at $H=20$ kOe is a sum of linear $T$ dependence and the $T^{2}$ dependence below $T_{SG}$. 

\begin{figure}
\includegraphics[width=8.0cm]{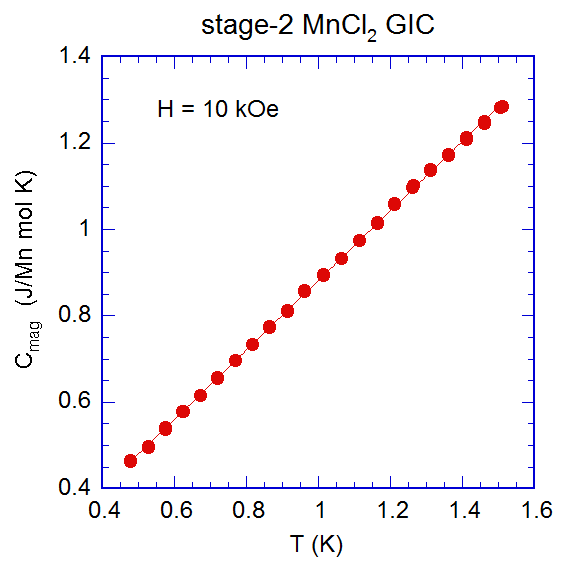}
\caption{\label{fig07}(Color online) Plot of $C_{mag}$ vs $T$ for $H = 10$ kOe. The solid line is a bets-fitting curve (straight line).}
\end{figure}

For $H = 10$ kOe, as shown in Fig.~\ref{fig07}, $C_{mag}$ is almost linearly proportional to $T$ at low temperatures. The least-squares fit of the data of $C_{mag}$ vs $T$ at $H = 10$ kOe in the temperature range between 0.45 and 1.5 K to 
\begin{equation} 
C_{mag} =\alpha_{1} +\delta_{1} T ,
\label{eq06}
\end{equation}
yields the parameters 
\[ 
\alpha_{1} = 0.077337 \pm 0.001986 \text{  J/Mn mol K}
\] 
and
\[ 
\delta_{1} = 0.80474 \pm 0.0019118 \text{  J/Mn mol K}^{2}.
\] 
We note that such a linear $T$ dependence of $C_{mag}$ at low temperatures is also observed in typical spin glasses such as Cu$_{0.988}$Mn$_{0.012}$\cite{ref18} and Eu$_{0.4}$Sr$_{0.6}$S.\cite{ref19} There are several models to explain the linear $T$ dependence in spin glass. These models are discussed in detail by Fisher and Herz.\cite{ref20} A $T^{2}$ dependence of $C_{mag}$ at low temperatures may indicates the presence of gapless and linearly dispersive modes in two dimensions. Such $T^{2}$ dependnce of $C_{mag}$ at low temperatures is observed in NiGa$_{2}$S$_{4}$.\cite{ref14} 

For $H = 0$, as shown in Fig.~\ref{fig05}, the upturn of the curve for $C_{mag}/T$ vs $T$ for $H = 0$ Oe is clearly seen in the low temperature ranges. Surprisingly, Wang and Swenson\cite{ref21} have predicted a peculiar property of the asymptotic low temperature specific heat, 
\begin{equation} 
C_{mag} =A\frac{1}{T^{2}}\exp (-\frac{\Delta }{T}) ,
\label{eq07}
\end{equation}
or
\begin{equation} 
\ln (C_{mag}T^{2})=\ln (A)-\frac{\Delta}{T} ,
\label{eq08}
\end{equation} 
using a replica Monte Carlo simulation method and finite-size transfer-matrix method for the 2D $\pm J$ Ising spin glass model, where $A$ is constant and $\Delta$ is an energy gap between the ground state and the first excited state in spin glass phase. The validity of this form has been confirmed by Saul and Kardar,\cite{ref21} Lukic et al.,\cite{ref23} and Katzbraber et al.,\cite{ref24} in spite of the difference in the physical interpretation of $\Delta$. The plot of $\ln(C_{mag}T^{2})$ vs $1/T$ should be a straight line with a slope $\Delta$. 

\begin{figure}
\includegraphics[width=8.0cm]{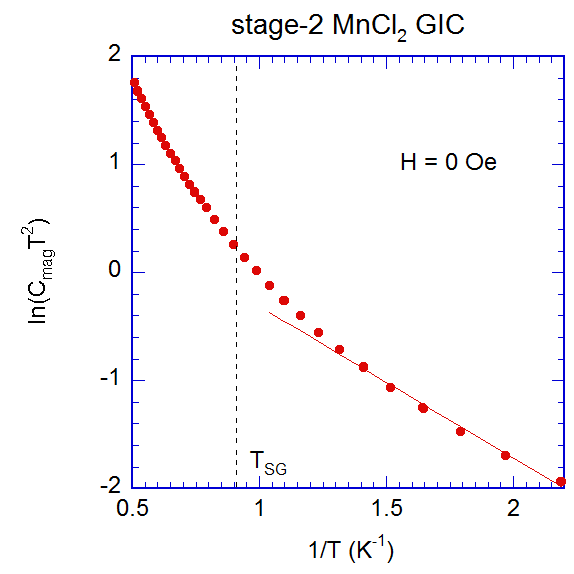}
\caption{\label{fig08}(Color online) Plot of $\ln(C_{mag}T^{2})$ vs $1/T$ at low temperatures for stage-2 MnCl$_{2}$ GIC. The straight line below $T_{SG}$ is a least-squares fitting curve denoted by Eq.(\ref{eq08}) with $\ln A = 1.0928$ and $\Delta = 1.407$ K.}
\end{figure}

Figures \ref{fig08} shows the plot of $\ln(C_{mag}T^{2})$ vs $1/T$ below $T_{SG}$ for stage-2 MnCl$_{2}$ GIC at $H = 0$. It is found that the curve of $\ln(C_{mag}T^{2})$ vs $1/T$ becomes straight line for $0.45<T<0.7$ K, as shown in Fig.\ref{fig08}. The least-squares fit of these data to Eq.(\ref{eq08}) yields the parameters 
\[
\Delta = 1.41 \pm 0.05 \text{  K,  }\ln(A) = 1.09 \pm 0.06.
\] 
The energy gap may disappear when the Zeeman energy ($g_{\perp}\mu_{B}SH$) as result of the application of $H$ to that system, becomes larger than the energy gap $k_{B}\Delta$, where $S$ = 5/2 and $g_{\perp} = 1.977$. The critical magnetic field $H_{0}$ is evaluated as 
\[
H_{0} \approx \frac{k_{B}\Delta}{g_{\perp}\mu_{B}S} = 4.25 \text{ kOe}.
\]
The value of $H_{0}$ is on the same order as the critical magnetic field at $T = 0$ K for the AT transition;\cite{ref17} $H_{AT} = 5.87 \pm  0.47$ kOe.

\subsection{\label{resultE}Entropy vs $T$}

\begin{figure}
\includegraphics[width=8.0cm]{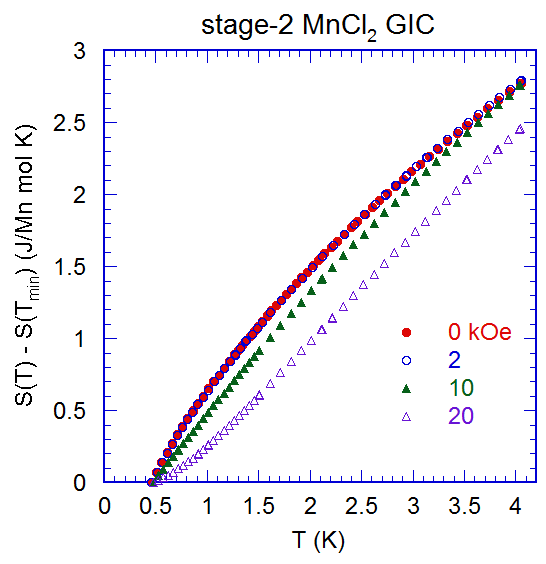}
\caption{\label{fig10}(Color online) Entropy $(= S(T)-S(T_{min})$ vs $T$, which is calculated from the data of $C_{max}$ vs $T$. $H$ = 0, 2 kOe, 10 kOe, and 10 kOe. $T_{min} = 0.45$ K.}
\end{figure}

\begin{figure}
\includegraphics[width=8.0cm]{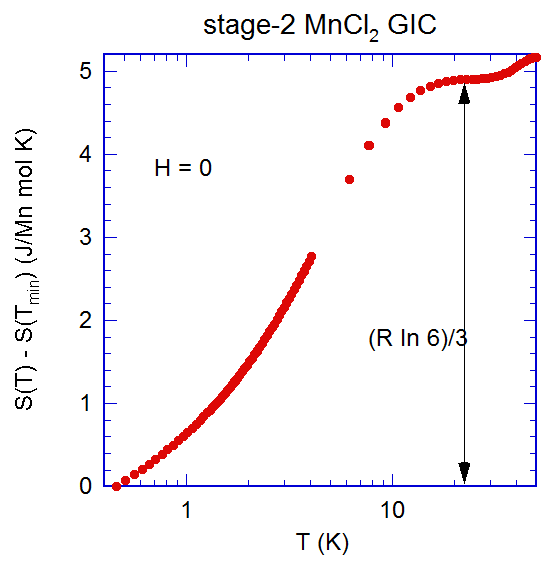}
\caption{\label{fig11}(Color online) Overview of entyropy $(= S(T)-S(T_{min}))$ vs $T$, which is calculated from the data of $C_{max}$ vs $T$. $H = 0$. $T_{min}$ = 0.45 K. The entropy due to the broad peak around 5 K is 1/3 of the total entropy.}
\end{figure}

Figure \ref{fig10} shows the plot of entropy $\Delta S=S(T)-S(T_{min})$ as a function of $T$ for stage-2 MnCl$_{2}$ GIC for $H$ = 0, 2, 10, 20 kOe, where $T_{min} = 0.45$ K. The entropy $\Delta S$ is defined as 
\begin{equation}
\Delta S=S(T)-S(T_{\min})=\int\limits_{T\min }^{T}\frac{C(T^{\prime})}{T^{\prime}} dT^{\prime} .
\label{eq09}
\end{equation}
Since $S = 5/2$, the total entropy $S_{total}$ is equal to
\[ 
S_{total}=R\ln (2S+1)=R\ln 6 = 14.898 \text{  J/Mn mol K}, 
\]
where $R$ is the gas constant and $R = 8.314472$ J/Mn mol K. As shown in Fig.~\ref{fig10}, the curve of $\Delta S$ vs $T$ at $H$ = 0 is upward convex, while the curve of $\Delta S$ vs $T$ at $H$ = 20 k Oe is downward convex. The curve of $\Delta S$ vs $T$ at $H$ = 0 is almost straight line, reflecting the linear $T$ dependence of $C_{mag}$ at $H$ = 10 kOe. Figure \ref{fig11} shows the $T$ dependence of $\Delta S$ at $H = 0$ in the wide temperature region up to 50 K. The entropy $\Delta S$ becomes flat between 20 and 30 K. The value of $\Delta S$ at 20 K is nearly equal to 1/3 of $R\ln 6$. This factor 1/3 is also observed in the entropy in NiGa$_{2}$S$_{4}$,\cite{ref14} which may be feature common to the 2D AFT. There are two step-like changes around 5 and 41 K corresponding to the peak temperatures of $C_{mag}$ vs $T$. It is considered that there is small amount of entropy contributing to the total entropy above 50 K, because of the absence of anomaly in $C_{mag}$ above 50 K. 

\begin{figure}
\includegraphics[width=8.0cm]{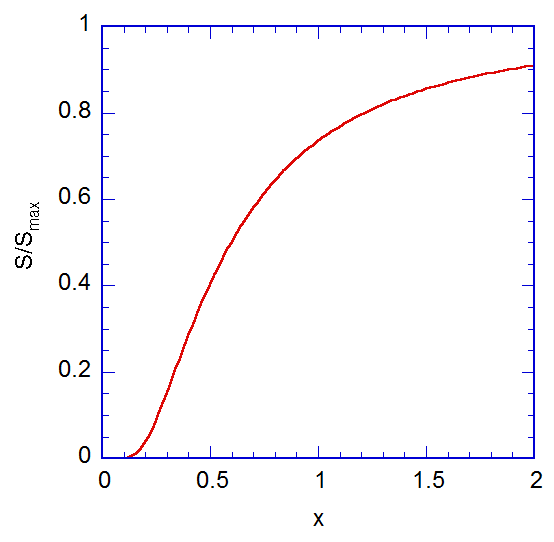}
\caption{\label{fig12}(Color online) Normalized entropy given by $S/S_{max}$ vs $x$ ($=T/\Delta$), where $S/S_{max}=f(x)$ defined by Eq.(\ref{eq11}). $\Delta $ = 1.41 K.}
\end{figure}

We now evaluate the entropy $S(T)$ at low temperatures. To this end, we assume that $C_{mag}$ is given by Eq.(\ref{eq07}). Then the entropy is calculated as 
\begin{eqnarray} 
S&=&\int\limits_{0}^{T}\frac{C}{T}dT=\int\limits_{0}^{T}\frac{A}{T^{3}}\exp(-\frac{\Delta }{T})dT \nonumber \\
&=&\frac{A}{T\Delta^{2}}(T+\Delta)\exp(-\frac{\Delta }{T}) ,
\label{eq10}
\end{eqnarray} 
leading to the scaling form of the entropy given by
\begin{equation} 
\frac{S}{S_{\max }}=f(x=\frac{T}{\Delta})=\frac{(1+x)\exp(-\frac{1}{x})}{x} ,
\label{eq11}
\end{equation}
with $x=T/\Delta$ and $S_{max}$ = $A/\Delta ^{2}$ [=1.507 J/(Mn mol K)], where $f(x)$ is a scaling function of $x$. In Fig.~\ref{fig12}, we show the plot of the normalized entropy $S/S_{max}$ as a function of $x$ ($=T/\Delta$). As is clearly shown, the curve of $S/S_{max}$ vs $x$ is upward convex, which is similar to the the experimental curve of $S$ vs $T$ at $H = 0$ (see Fig.~\ref{fig10}). 

When $T_{min}$ = 0.45 K and $\Delta$ = 1.41 K, $S(T_{min})$ can be evaluated as $S(T_{min})= 1.507 f(x=T_{min}/\Delta)=0.272$ J/Mn mol K from Fig.~\ref{fig12}. Then the observed entropy at 50 K is calculated as 
\[
S(T=50 \text{ K})=S(T_{min})+\Delta S= 5.47 \text{  J/Mn mol K}
\] 
This value of $S(T = 50 K)$ is much smaller than the total entropy (= 14.898 J/Mn mol K). The missing entropy may be a residual entropy at $T = 0$ K because of highly frustrated nature of the system. This residual entropy is calculated as 
\[
S_{residual}\approx S_{total} - S(T=50\text{ K}) = 9.43 \text{  J/Mn mol K}
\] 
which corresponds to 63 \% of the total entropy. The residual entropy occurs when the system can exist in a large number of ground states with the same zero-point energy. Note that for highly frustrated pyrochlore, Dy$_{2}$Ti$_{2}$O$_{7}$,\cite{ref25} the residual entropy is 1/3 of the total entropy. Here we note that even the residual entropy is finite, still the magnetic specific heat is zero at $T$ = 0 K,\cite{ref26} since 
\begin{eqnarray}
S_{residual} &=& \lim_{T\rightarrow 0}S=\lim_{T\rightarrow 0}\frac{TS}{T}=\lim_{T\rightarrow 0}(T\frac{dS}{dT}+S) \nonumber\\
\lim_{T\rightarrow 0}C &=& \lim_{T\rightarrow 0}T\frac{dS}{dT}=0 ,
\label{eq12}
\end{eqnarray}

\section{\label{sum}Summary of results from previous works and further analysis}
\subsection{\label{sumA}Magnetic neutron scattering}

\begin{figure}
\includegraphics[width=8.0cm]{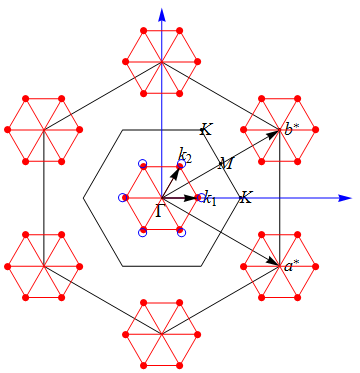}
\caption{\label{fig13}(Color online) Magnetic Bragg reflections (denoted by solid circles) in the in-plane reciprocal lattice vector. $\left| {\bf a}^{*}\right|=\left| {\bf b}^{*}\right| = 1.965 \AA^{-1}$. $\left| {\bf k}_{1}\right|=\left| {\bf k}_{2}\right| = 0.522 \AA^{-1}$. The $K$ point is denoted by $({\bf a}^{*} +{\bf b}^{*})/3$ or $120^\circ$ spin structure or $\sqrt{3}\times\sqrt{3}$ spin structure; $\left| {\bf Q}_{com}\right|=\sqrt{3} \left| a^{*} \right| /3=1.1345 \AA^{-1}$. The points denoted by open circles correspond to the commensurate structure $2\sqrt{3}\times 2\sqrt{3}$ spin structure. These points are expressed by $({\bf a}^{*} +{\bf b}^{*})/6$; $\left| {\bf Q}_{com}\right|=\sqrt{3}\left| {\bf a}^{*}\right| /6=0.5672 \AA^{-1}$. Note that the location of magnetic reflections is close to, but not equal to the magnetic Bragg reflections corresponding to the $2\sqrt{3}\times 2\sqrt{3}$ commensurate spin structure. The reciprocal lattice of the graphite layer is not included in this figure.}
\end{figure}

The MnCl$_{2}$ layer forms a triangular Mn$^{2+}$ lattice in stage-2 MnCl$_{2}$-GIC. Primitive lattice vectors {\bf a} and {\bf b} have length $\left| {\bf a}\right|=\left| {\bf b}\right| = 3.692 \pm 0.005 \AA$, the same as for pristine MnCl$_{2}$ ($3.693 \AA$). The in-plane reciprocal lattice vector is given by ${\bf a}^{*}$ and ${\bf b}^{*}$ where $\left| {\bf a}^{*}\right|=\left| {\bf b}^{*}\right| = 4\pi/(\sqrt{3} a) = 1.965 \AA^{-1}$. The magnetic neutron scattering results of stage-2 MnCl$_{2}$ GIC have been reported previously by Wiesler et al.\cite{ref08,ref09,ref10} The in-plane magnetic scattering at 0.43 K is obtained by subtracting the intensity at 14.85 K from the corresponding intensity at 0.43 K. Magnetic peaks are observed at $\left| {\bf Q}_{\perp}\right|$ = 0.522, 1.536, 2.05, and 2.44 $\AA^{-1}$ (see Figs. 5 and 6 of Ref. \cite{ref10}), where ${\bf Q}_{\perp}$ is the in-plane wave vector. These wave vectors are consistent with the model shown in Fig.~\ref{fig13}. According to this model, the magnetic Bragg reflections should appear at the in-plane wave vectors ${\bf Q}_{\perp} = {\bf G}(h,k) \pm {\bf k}_{1}$, ${\bf G}(h, k) \pm  {\bf k}_{2}$, or ${\bf G}(h, k) \pm ({\bf k}_{1} - {\bf k}_{2})$, where ${\bf G}(h, k) = h {\bf a}^{*} + k {\bf b}^{*}$ is the in-plane reciprocal lattice vector of MnCl$_{2}$ layer, $h$ and $k$ are integers, and ${\bf k}_{i}$ ($i$ = 1, 2) are the reciprocal lattice vectors of magnetic superlattice. Assigning $\left| {\bf k}_{1}\right| = 0.522 \AA^{-1}$ and $\left| {\bf a}^{*} - {\bf k}_{1}\right| =1.536 \AA^{-1}$, the angle $\theta$ between ${\bf k}_{1}$ and ${\bf a}^{*}$ can be estimated as 
\begin{equation}
\cos\theta=\frac{\left| {\bf a}^{*}\right|^{2}+\left| {\bf k}_{1}\right|^{2}-\left| {\bf a}^{*}-{\bf k}_{1}\right|^{2}}{2\left| {\bf k}_{1}\right| \left| {\bf a}^{*}\right|} ,
\label{eq13}
\end{equation} 
or $\theta = 30 \pm 2^\circ$. The magnetic Bragg reflections are predicted to occur at $\left| {\bf Q}_{\perp}\right|=\left| {\bf a}^{*}+{\bf k}_{2}\right| = 2.03 \AA^{-1}$, and $\left| {\bf a}^{*}+{\bf k}_{1}\right| = 2.431 \AA^{-1}$, in good agreement with the experimental values, $\left| {\bf Q}_{\perp}\right|$ = 2.05 and 2.44 $\AA^{-1}$, respectively. 

\begin{figure}
\includegraphics[width=8.0cm]{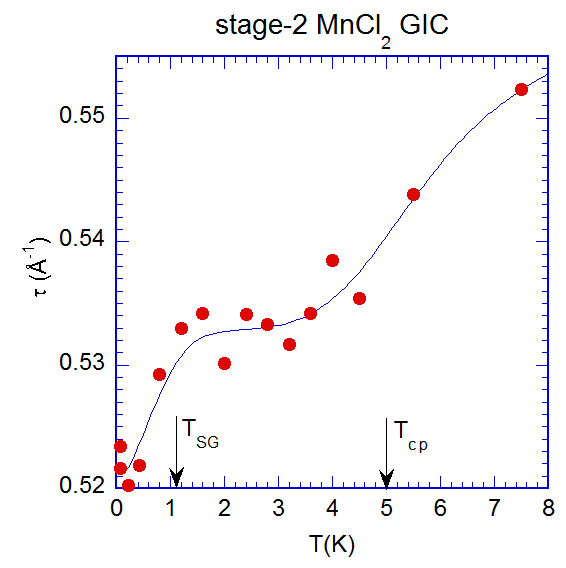}
\caption{\label{fig14}(Color online) $T$ dependence of the center of the Lorenzian peak, $\tau$, for stage-2 MnCl$_{2}$ GIC. The original data of $\tau$ vs $T$ were taken by Wiesler et al.\cite{ref10} $\tau=\left| {\bf Q}_{com}\right|=\sqrt{3} \left| a^{*} \right| /6=0.5672 \AA^{-1}$ corresponds to the commensurate in-plane spin structure with the periodicity $2\sqrt{3}\times 2\sqrt{3}$. The solid line is a guide to the eyes. $T_{SG}=1.1$ K. $T_{cp}\simeq 5$ K.}
\end{figure}

The temperature dependence of the scattering has been also studied by a series of in-plane scans at temperatures between 63 mK and 7.5 K. Throughout the entire temperature range, the intensities fit well to a Lorentzian peak described by 
\begin{equation}
I=\frac{I_{0}\kappa}{\pi \lbrack (\left| Q_{\perp}\right|-\tau)^{2}+\kappa^{2}\rbrack} ,
\label{eq14}
\end{equation} 
around $\left| {\bf Q}_{\perp}\right|=\left| {\bf k}_{1}\right|$, where $\kappa$ is the inverse in-plane spin correlation length, $\tau$ is the peak position, and $I_{0}$ is the integrated intensity. In Fig.~\ref{fig14} we show the $T$ dependence of the peak position $\tau$. The peak position $\tau$ is nearly constant at $0.522 \AA^{-1}$ below 0.45 K and increases rapidly with increasing $T$ in the vicinity of $T_{SG}$ (= 1.1 K). Above $T_{SG}$, $\tau$ plateaus briefly at $0.532 \AA^{-1}$, rising again above 4.5 K (close to the peak temperature $T_{cp}$ of $C_{mag}$ vs $T$) toward the commensurate position at $\left| {\bf Q}_{com}\right|=\sqrt{3} \left| {\bf a}^{*} \right| /6=0.5672 \AA^{-1}$ (see the original data of $\tau$ vs $T$ in Fig. 7(a) in Ref. \cite{ref10}), where ${\bf Q}_{com} = ({\bf a}^{*} +{\bf b}^{*})/6$. 

\begin{figure}
\includegraphics[width=8.0cm]{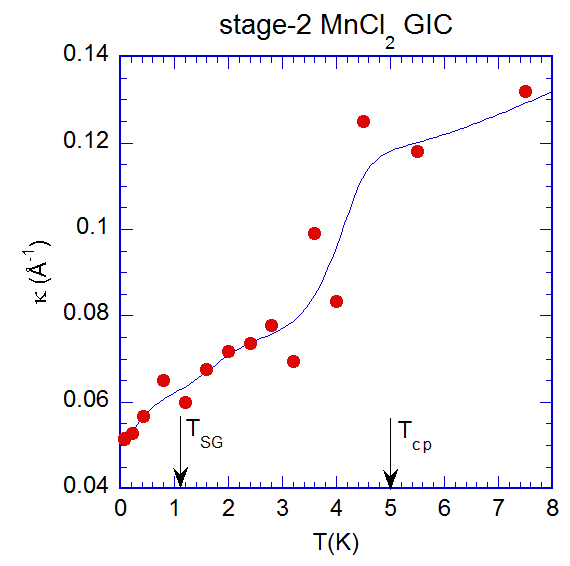}
\caption{\label{fig15}(Color online) Inverse in-plane correlation length $\kappa$ ($=1/\xi$) vs $T$. $\xi$ is the in-plane spin correlation length. The instrumental resolution limit for $\kappa$ is $0.016 \AA^{-1}$. The original data of $\kappa$ vs $T$ were taken by Wiesler et al.\cite{ref10} The solid line is a guide to the eyes. $T_{SG}=1.1$ K. $T_{cp}\simeq 5$ K.}
\end{figure}

In Fig.~\ref{fig15} we show the $T$ dependence of the inverse in-plane spin correlation length $\kappa$ (see also the original data of $\kappa$ vs $T$ in Fig. 7(c) in Ref. \cite{ref10}). The value of $\kappa$ is $0.0566 \AA^{-1}$ at 0.45 K, corresponding to an in-plane spin correlation length $\xi$ ($=1/\kappa$) of only $18 \AA$. The value of $\kappa$ slowly increases with increasing $T$. It does not show any anomaly at $T_{SG}$. It undergoes a step-like change around 4.5 K, at which $C_{mag}$ shows a broad peak. The value of $\kappa$ is $0.124 \AA^{-1}$ at 4.5 K, corresponding to an in-plane spin correlation length $\xi$ of only $8 \AA$.

\subsection{\label{sumB}ZFC and FC magnetization}

\begin{figure}
\includegraphics[width=8.0cm]{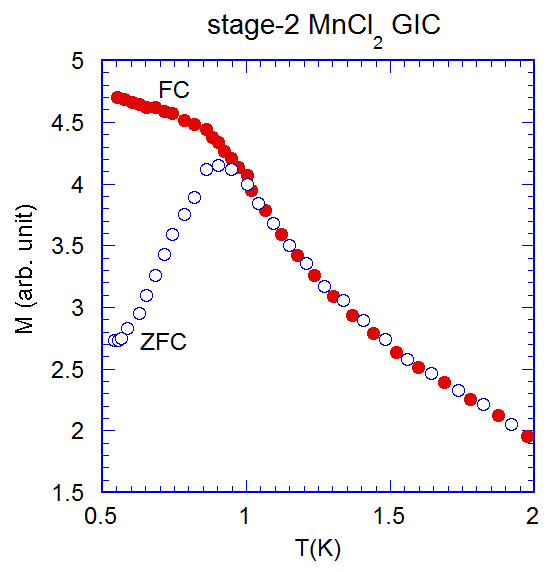}
\caption{\label{fig16}(Color online) $T$ dependence of $M_{ZFC}$ an $M_{FC}$ for stage-2 MnCl$_{2}$ GIC. See the detail of ZFC and FC cooling protocol in the text. The magnetization is measured at $H$ = 100 mOe. The original data were taken by Matsuura et al.\cite{ref06}}
\end{figure}
 
The $T$ dependence of zero-field cooled (ZFC) magnetization and field-cooled (FC) magnetization for stage-2 MnCl$_{2}$ GIC has been reported by Matsuura et al.\cite{ref06} Their result is reproduced in Fig.~\ref{fig16}. In the ZFC cooling protocol, the system is quenched from high temperatures well above $T_{SG}$ down to 0.5 K in the absence of an external magnetic field. Then the external magnetic field ($H$ = 100 mOe) is applied to the system at $T$ = 0.5 K. The ZFC magnetization is measured with increasing $T$ from 0.5 K to temperatures well above $T_{SG}$ under the presence of $H$. The FC magnetization is measured with decreasing $T$ from high temperatures well above $T_{SG}$ down to 0.5 K, in the presence of $H$ (= 100 mOe). The $T$ dependence of $M_{ZFC}$ and $M_{FC}$ is very similar to that of typical spin glases. $M_{ZFC}$ shows a peak around $T$ = 0.9 K. $M_{ZFC}$ starts to deviate from $M_{FC}$ below $T_{SG}$ ($\approx 1.0$ K). In other words, the irreversible effect of magnetization occurs below $T_{SG}$. The value of $T_{SG}$ thus determined from Fig.~\ref{fig16} is a little different from that of the AC magnetic susceptibility reported by Kimishima et al.\cite{ref03,ref04} Here we define that $T_{SG}$ is equal to 1.1 K. The magnetization $M_{FC}$ bends, becoming almost flat just below $T_{SG}$, which is one of the significant features of spin glasses. In summary, the stage-2 MnCl$_{2}$ GIC undergoes a spin glass phase transition at $T=T_{SG}$. 

\subsection{\label{sumC}AC magnetic susceptibility, de Almeida-Thouless line}

\begin{figure}
\includegraphics[width=8.0cm]{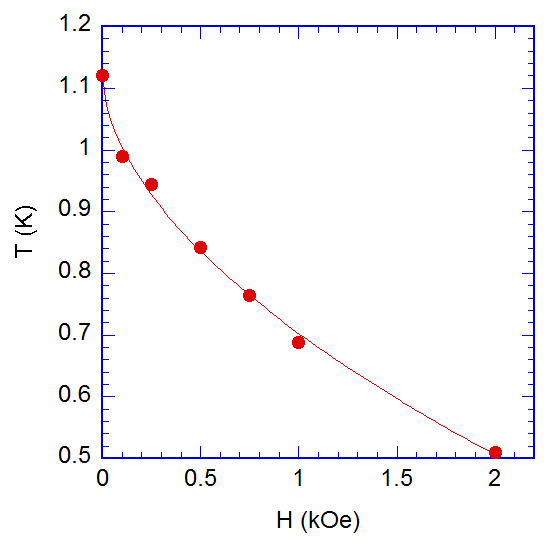}
\caption{\label{fig17}(Color online) Plot of the peak temperature ($\chi^{\prime}$ vs $T$ in the presence of $H$) as a function of $H$ for stage-2 MnCl$_{2}$ GIC. $H\parallel c$ plane. The original data were taken by Kimishima et al.\cite{ref03,ref04} The solid line is a best-fitting curve to Eq.(\ref{eq15}).}
\end{figure}

Kimishima et al.\cite{ref03,ref04} have reported the $T$ dependence of AC magnetic susceptibility (the dispersion $\chi^{\prime}$) for stage 2 MnCl$_{2}$ GIC in the presence of an external magnetic field $H$ along the $c$ plane. The dispersion $\chi^{\prime}$ shows a peak at $T=T_{SG}$ (= 1.1 K). This peak shifts to the low temperature side with increasing $H$. In Fig.~\ref{fig17}, we show the $H$ dependence of the peak temperature for stage-2 MnCl$_{2}$ GIC which has been previously reported by Kimishima et al.\cite{ref04} It is predicted for typical spin glasses that the peak temperature $T_{SG}(H)$ forms a de-Almeida-Thouless (AT)\cite{ref17} line in the $H$-$T$ magnetic phase diagram, which is defined by
\begin{equation} 
H=H_{AT} \lbrack 1-\frac{T_{SG}(H)}{T_{SG}H=0)} \rbrack ^{p} ,
\label{eq15}
\end{equation} 
where $p$ is an exponent and $p$ = 3/2, and $H_{AT}$ is the magnetic field at $T$ = 0 K. The least-squares-fit of the data of $T_{SG}(H)$ vs $H$ to Eq.(\ref{eq15}) yields the parameters 
\[
T_{SG}(H=0) = 1.184 \pm 0.013 \text{ K}, 
\]
\[
p = 1.79 \pm 0.10,\text{ and } H_{AT} = 5.87 \pm 0.47 \text{kOe}.
\] 
The exponent $p$ thus obtained is relatively close to $p$ = 3/2 for the AT line. This indicates that the low temperature phase below $T_{SG}$ is a spin glass phase. Here we note that the characteristic field $H_{AT}$ is theoretically predicted as (Katori and Ito)\cite{ref27} 
\begin{equation}
H_{AT}=\sqrt{\frac{8}{(m+1)(m+2)}}\frac{k_{B}T_{SG}(H=0)}{g\mu_{B}S} ,
\label{eq16}
\end{equation} 
where $m=1$ for the Ising symmetry and $m=3$ for the Heisenberg symmetry, $g$ is the Land\'{e} $g$-factor, $k_{B}$ is the Boltzmann constant, and $\mu_{B}$ is the Bohr magneton. The value of $H_{AT}$ is evaluated as 2.103 kOe for $m=1$ and as 3.839 kOe for $m=3$, where $g=g_{\perp}=1.97$ for stage-2 MnCl$_{2}$ GIC. These values of $H_{AT}$ is smaller than tha derived from the analysis of the least-squares fitting. 

\subsection{\label{sumD}DC magnetic susceptibility}

\begin{table*}
\caption{\label{Table1}Parameters of the structure and magnetism in stage-2 MnCl$_{2}$ GIC.\cite{ref03,ref04,ref05,ref06,ref07,ref08,ref09,ref10}}
\begin{ruledtabular}
\begin{tabular}{ccl}
parameter & value & description\\
\hline
$\left| {\bf a}^{*}\right|=\left| {\bf b}^{*}\right|$ & 1.965 $\AA^{-1}$ & in-plane reciprocal lattice of MnCl$_{2}$ layer\\
$\left| {\bf Q}_{comm}\right|$ & 0.5672 $\AA^{-1}$ & wave number for the $2\sqrt{3}\times 2\sqrt{3} $ 
 spin structure\\
$\left| {\bf Q}_{incomm}\right|$ & 0.522 $\AA^{-1}$ & wave number for the incommensurate short range spin order at 0.45 K\\
$a$ & 3.692 $\AA$ & in-plane lattice constant of MnCl$_{2}$ layer\\
$d$ & 12.80 $\AA$ & $c$-axis repeat distance\\
$\Theta_{\perp}$ & - 9.1 K & Curie-Weiss temperature along the $c$ plane\\
$\Theta_{\parallel}$ & - 5.9 K & Curie-Weiss temperature along the $c$ axis\\
$P_{\perp}$ & 5.831 $\mu_{B}$ & effective magnetic moment along the $c$ plane\\
$P_{\parallel}$ & 5.838 $\mu_{B}$ & effective magnetic moment along the $c$ axis\\
$g_{\perp}$ & 1.977 & $g$-factor along the $c$ plane at room temperature\\
$g_{\parallel}$ & 1.912 & $g$-factor along the $c$ axis at room temperature\\ 
$J_{1}$ & - 0.20 K & antiferromagnetic n.n. intraplanar exchange interaction\\ 
$D$ & 0.97 K & positive single ion anisotropy, showing $XY$ symmetry\\
$T_{SG}$  & 1.1 K & spin freezing temperature\\
$S$ & 5/2 & spin of Mn$^{2+}$\\
\end{tabular}
\end{ruledtabular}
\end{table*}

The DC magnetic susceptibility of stage-2 MnCl$_{2}$ GIC has been reported by Wiesler et al.\cite{ref05} The parameters of the magnetic properties are listed in Table \ref{Table1}. Both $\chi_{\parallel}$ and $\chi_{\perp}$ (measured in the FC state) decrease monotonically with increasing $T$ and exhibits no anomaly around 41 K, where the magnetic specific heat $C_{mag}$ shows a broad peak. $\chi_{\perp}>\chi_{\parallel}$ at all temperatures, which suggests that spins align in the MnCl$_{2}$ layer. The anisotropy $\Delta\chi=\chi_{\perp}-\chi_{\parallel}$ becomes appreciable only below 50 K, a result with a shift of the $g$-values below 50 K, may indicate that onset of short-range spin order with {\bf Q} = 0 mode in the MnCl$_{2}$ layer. 

As listed in Table \ref{Table1}, the Curie-Weiss temperature $\Theta_{\perp}$ along the $c$ plane is equal to $-9.1$ K, and the Curie-Weiss temperature $\Theta_{\parallel}$ along the $c$ axis is equal to $-5.9$ K. The negative sign of $\Theta_{\perp}$ and $\Theta_{\parallel}$ indicates that the intraplanar exchange interaction is antiferromagnetic. The peak temperature ($T_{cp}\simeq 5$ K) of $C_{mag}$ vs $T$ is on the same order as $\left|\Theta_{\parallel}\right|$ and $\left|\Theta_{\perp}\right|$. 

\subsection{\label{sumE}Line width of electron spin resonance}

\begin{figure}
\includegraphics[width=8.0cm]{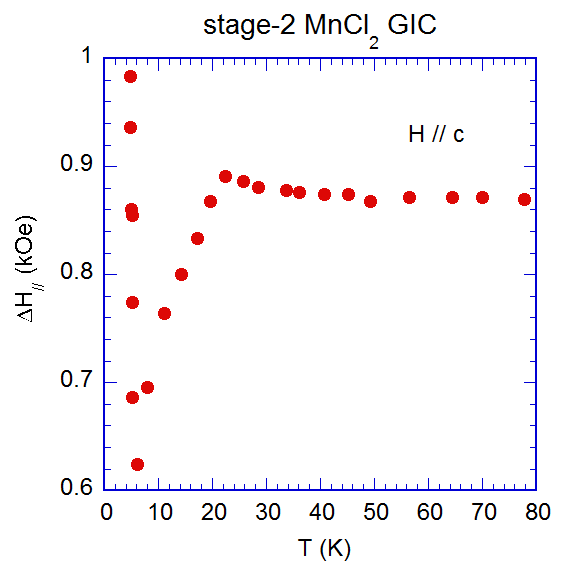}
\caption{\label{fig18}(Color online) $T$ dependence of the ESR line-width $\Delta H$ for stage-2 MnCl$_{2}$ GIC. $H\parallel c$. The ESR measurement was made by the conventional field-modulated X-band spectrometer at frequency $\nu$ = 9.42 GHz in the temperature range between 1.38 and 293 K. The original data were taken by Koga and Suzuki,\cite{ref02} Suzuki et al.\cite{ref08}}
\end{figure}

The electron spin resonance (ESR) of stage-2 MnCl$_{2}$ GIC has been measured by Koga and Suzuki,\cite{ref02} and Suzuki et al.\cite{ref07} Figure \ref{fig18} shows the $T$ dependence of ESR linewidth $\Delta H$ for stage-2 MnCl$_{2}$ GIC when the external magnetic field $H$ is applied along the $c$ axis ($\parallel$). The line width $\Delta H$ is defined as the peak-to-peak width of the derivative ESR line shape.

The $T$ dependence of $\Delta H_{\parallel}$ for $H\parallel c$ is consistent with that of $T\chi_{\parallel}$ for $6\le T\le 20$ K. This suggests that the {\bf Q} = 0 mode is dominant in these temperature ranges. As $T$ decreases from 6 K ($\approx\left|\Theta_{\parallel}\right|$ = 5.9 K), the mode ${\bf Q} = {\bf Q}_{incomm}$ of the spin fluctuations takes over the {\bf Q} = 0 mode, where {\bf Q}$_{incomm}$ is the wave vector corresponding to the in-plane incommensurate spin structure (close to the $2\sqrt{3} \times 2\sqrt{3}$ spin structure). The minimum of $\Delta H_{\parallel}$ around 6 K results from the competition between the {\bf Q} = 0 mode and ${\bf Q} = {\bf Q}_{incomm}$ mode. The magnitude $\tau$ (= $\left| {\bf Q}_{incomm}\right|$) decreases with decreasing $T$ as is observed from the magnetic neutron measurements. The linewidth $\Delta H_{\parallel}$ diverges as $T$ approaches $T_{SG}$ from the high temperature side. 

\section{\label{dis}Discussion}
\subsection{\label{disA}Nature of short range spin order in the vicinity of $T_{SG}$}

\begin{figure}
\includegraphics[width=8.0cm]{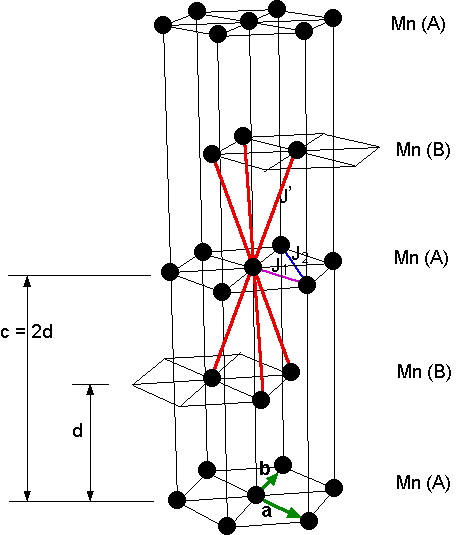}
\caption{\label{fig19}(Color online) Simplified model for the hexagonal-closed packed (hcp) type structure of stage-2 MnCl$_{2}$ GIC, where only the positions of Mn$^{2+}$ ions are shown. $a = 3.692 \AA$. $d = 12.80 \AA$. {\bf a}, {\bf b} and {\bf c} are lattice vectors.}
\end{figure}

The origin of the incommensurate in-plane short range spin order is discussed, based on the model proposed by Sakakibara\cite{ref28} who explains the magnetic structure of C$_{6}$Eu (donor-type stage-1 GIC), where C$_{6}$Eu has a hexagonal closed packed (hcp) type structure. Here we assume that stage-2 MnCl$_{2}$ GIC also have the hcp type structure, which is schematically shown in Fig.~\ref{fig19}. For simplicity, only the positions of Mn$^{2+}$ ions are shown. From the symmetry of the stacking sequence in hcp-type structure, we assume that there are two kinds of stacking sequences such as $ABABAB$ (Fig.~\ref{fig19}) and $ACACAC\cdots$ in the weight of 1 to 1, where $A$, $B$, and $C$ layers forms a triangular lattice. The positions of Mn$^{2+}$ ions in the $B$ and $C$ layers are shifted by the translation vectors ${\bf \Delta}$ and $2{\bf \Delta}$ with respect to those of atoms in the $A$ layer, where 
\[
{\bf \Delta}=\frac{1}{3} \left( 2{\bf a}+{\bf b} \right) .
\] 
The lattice vectors are described in terms of the Cartesian coordinates, as
\[ 
{\bf a}=\left( \frac{1}{2} a,-\frac{\sqrt{3} }{2} a,0\right)  ,
{\bf b}=\left( \frac{1}{2} a,\frac{\sqrt{3} }{2} a,0\right)  ,
{\bf c}=\left( 0,0,2d\right) .
\]
where $a$ is the lattice constant (= $3.692 \AA$, $d$ is the $c$-axis repeat distance ($d = 12.80 \AA$), and $c=2d$ is the lattice constant along the $c$ axis. The reciprocal lattice vectors are also described in terms of the Cartesian coordinates, as
\[ 
{\bf a}^{*} =\frac{4\pi }{\sqrt{3} a} \left( \frac{\sqrt{3} }{2} ,-\frac{1}{2} ,0\right)  ,
{\bf b}^{*} =\frac{4\pi }{\sqrt{3} a} \left( \frac{\sqrt{3} }{2} ,\frac{1}{2} ,0\right)  ,
\]
\[
{\bf c}^{*} =\frac{2\pi }{2d} \left( 0,0,1\right) .
\]
where $a^{*} =\left| {\bf a}^{*} \right| =4\pi /(\sqrt{3} a)$ and $c^{*} =\left| {\bf c}^{*} \right| =\pi /d$ . The $K$ point in the Brillouin zone (see Fig.~\ref{fig13}) is expressed by 
\[
\frac{1}{3} ({\bf a}^{*} +{\bf b}^{*} )=\frac{4\pi }{3a} \left( 1,0,0\right) =a^{*} \frac{1}{\sqrt{3} } \left( 1,0,0\right) .
\]
The reciprocal vectors ${\bf a}^{*}$ and ${\bf b}^{*}$ are denoted in Fig.~\ref{fig13}, where the $K$ point is on the $Q_{x}$ axis. According to Sakakibara,\cite{ref28} the minimum energy of the system is given by 
\begin{equation}
E_{\min }=-NS^{2}J({\bf Q}),
\label{eq17}
\end{equation} 
with
\begin{equation} 
J({\bf Q})=\sum\limits_{{\bf R}_{ij}}J({\bf R}_{ij})e^{i{\bf Q}\cdot {\bf R}_{ij}}  ,
\label{eq18}
\end{equation} 
where {\bf Q} is the characteristic wave vector, the $N$ is total number of Mn$^{2+}$ ions, ${\bf S}_{i}$ is the classical spin vector of Mn$^{2+}$ ion at the ${\bf R}_{i}$ site, ${\bf R}_{ij} ={\bf R}_{i} -{\bf R}_{j}$, $J({\bf R}_{ij})$ is the exchange interaction between ${\bf S}_{i}$ at the ${\bf R}_{i}$ site and ${\bf S}_{j}$ at ${\bf R}_{j}$ site: $J({\bf R}_{ij}) = J(-{\bf R}_{ij})$. In our system, $J({\bf Q})$ is defined by 
\begin{equation}
J({\bf Q})\equiv J_{1}({\bf Q})+J_{2}({\bf Q})+J_{3}({\bf Q})+J^{\prime}({\bf Q}) ,
\label{eq19}
\end{equation}
with 
\begin{eqnarray}
J_{1} ({\bf Q}) &=& 2J_{1} \lbrack \cos (aQ_{x} )+2\cos (\frac{aQ_{x}}{2} )\cos (\frac{\sqrt{3} }{2} aQ_{y} )\rbrack , \label{eq20}\\
J_{2} ({\bf Q}) &=& 2J_{2} \lbrack \cos (\sqrt{3} aQ_{y} )+2\cos (\frac{3}{2} aQ_{x} )\cos (\frac{\sqrt{3} }{2} aQ_{y} )\rbrack,\nonumber \\\label{eq21}\\
J_{3} ({\bf Q}) &=& 2J_{3} \lbrack \cos (2aQ_{x} )+2\cos (aQ_{x} )\cos (\sqrt{3} aQ_{y} )\rbrack , \label{eq22}\\ 
J_{4} ({\bf Q}) &=& 2J^{\prime}\cos (dQ_{z} )\lbrack 2\cos (\frac{aQ_{x} }{2} )\cos (\frac{aQ_{y} }{2\sqrt{3} } )+\cos (\frac{aQ_{y} }{\sqrt{3} } )\rbrack,\nonumber \\ \label{eq23}
\end{eqnarray} 
where $J_{1}$ is the nearest neighbor (n.n.) intraplanar exchange interaction (antiferromagnetic in the present system), $J_{2}$ is the next-nearest neighbor (n.n.n.) intrapnar exchange interaction, $J_{3}$ is the next next-nearest neighbor (n.n.n.n.) intrapnar exchange interaction, and $J^{\prime}$ is the interplanar exchange interaction. The wavenumbers $Q_{x}$, $Q_{y}$, and $Q_{z}$ are defined as 
\[
Q_{x} =\frac{4\pi }{\sqrt{3} a} h=a^{*} h , 
Q_{y} =\frac{4\pi }{\sqrt{3} a} k=a^{*} k ,
Q_{z} =\frac{\pi }{d} \zeta =c^{*} \zeta  . 
\]
The in-plane spin structure is realized as a result of the minimum energy state with a characteristic in-plane wave vector {\bf Q}, as 
\begin{equation}
{\bf S}_{i}=S\cos({\bf Q}\cdot {\bf R}_{i}){\bf e}_{x} +S\sin({\bf Q}\cdot {\bf R}_{i} ){\bf e}_{y} ,
\label{eq24}
\end{equation} 
on the triangular lattice site ${\bf R}_{i}$,
 
We need to find the location of the magnetic Bragg points in the {\bf Q} space for each pair of $J_{1}$, $J_{2}$, $J_{3}$, and $J^{\prime}$. The magnetic Bragg reflection appears at the vector {\bf Q}, where $J$({\bf Q}) has a maximum. To this end, we use a ContourPlot (equi-energy contour plot) of $J({\bf Q})$ in the Mathematica, where the points with the same $J({\bf Q})$ are connected, forming an contour line. When $J_{1} = -1$ (for simplicity the magnitude is normalized as 1, antiferromagnetic), $J_{2}$ = $J_{3}$ = $J^{\prime}$ = 0, the magnetic Bragg points appear at the K point, $(1/3 {\bf a}^{*} + 1/3 {\bf b}^{*})$. The in-plane spin structure with {\bf Q} at the $K$ point is a $120^\circ$ spin structure ($\sqrt{3} \times \sqrt{3}$) as is expected. 

\begin{figure}
\includegraphics[height=4.5cm]{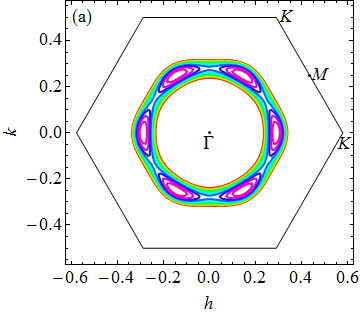} 
\includegraphics[height=4.5cm]{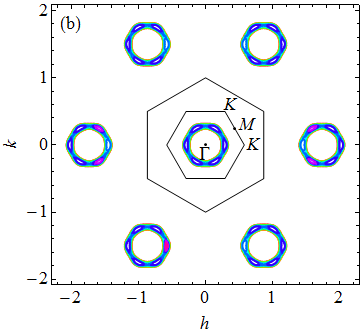}
\caption{(Color online) Energy contour line of $J({\bf Q})$ for $J_{1} = -1$, $J_{2}$ = 0, $J_{3}$ = 0, and $J^{\prime}$ = 2 in the $(h, k)$ plane. $Q_{x}=a^{*}h$. $Q_{y}=a^{*}k$. $Q_{z} = 0$. $h$ = 0.28675 and $k$ = 0 for the $\left| {\bf Q}_{comm}\right| = 0.5672 \AA^{-1}$.}
\label{fig20}
\end{figure}

We determine the magnetic phase diagram in the ($J_{2}$, $J_{4}$) plane where $J_{1}$ = -1 and $J_{3}$ = 0. The magnetic phase diagram consists of the K point reflection ($\sqrt{3}\times\sqrt{3}$), the M point reflection ($2 \times 2$), and ($\sqrt{7}\times\sqrt{7}$) reflections, depending on the values of $J_{2}$ and $J_{4}$. The detail of the magnetic phase diagram will be reported elsewhere. In the limited case of $J_{1}$ = -1, $-0.02\le J_{2}\le 0.08$, and $J^{\prime}$ = 2, the magnetic Bragg points appear at the point $({\bf a}^{*}/6 + {\bf b}^{*}/6)$, or $Q_{x}=ha^{*}$ and $Q_{y}=k$ $a^{*}$ with $h$ = $\sqrt{3} /6$ = 0.28675 and $k$ = 0. In Figs.~\ref{fig20}(a) and (b), we show the energy-contour plot of $J({\bf Q})$, where $J_{1} = -1$, $J_{2}$ = 0 , and $J^{\prime}$ = 2. It is clear that $J({\bf Q})$ has a maximum at $h$ = 0.28675 and $k$ = 0, which is the Bragg peak from the commensurate in-plane spin structure. Note that the energy contour plot of Fig.~\ref{fig20}(b) is similar to the schematic diagram of the reciprocal lattice phase shown in Fig.~\ref{fig13}. Because of $J^{\prime}>0$, $J^{\prime}$ is a ferromagnetic interplanar interaction. The magnitude of $J^{\prime}$ is larger than that of $J_{1}$. This result seems to be inconsistent with our assumption that $\left| J^{\prime}\right|$ is much smaller than $\left| J_{1}\right|$. However, if the effective interplanar exchange interaction $J_{eff}^{\prime}$ is used instead of $J^{\prime}$ and the magnitude of $J_{eff}^{\prime}$ becomes large with decreasing $T$, the above inconsistency may be ruled out. Below 30 K where the static magnetic susceptibility becomes anisotropic, the short-range spin order gradually grows with decreasing $T$. This growth leads to the effective interplanar exchange interaction between adjacent MnCl$_{2}$ layers, 
\begin{equation}
J^{\prime}_{eff} =J^{\prime}(\xi_{\perp}/a)^{2} ,
\end{equation} 
with $\xi_{\perp} = 18 \AA$ at 0.45 K and $a = 3.692 \AA$. Even if $\left| J^{\prime}\right|$ is very small, the effective interaction $\left| J^{\prime}_{eff}\right|$ becomes large when $\xi_{\perp}$ become large with decreasing $T$. Then $\left| J^{\prime}_{eff}\right|$ gives rise to a spin correlation between Mn$^{2+}$ ions of adjacent MnCl$_{2}$ layers. 

In the present stage, we cannot find any appropriate values of ($J_{2}$, $J_{4}$) for the magnetic Bragg peak at the incommensurate spin structure ($h$ = 0.2656, $k$ = 0) corresponding to $\left| {\bf Q}_{incomm}\right| = 0.522 \AA^{-1}$ from our method. Note that this situation may change when $J_{3}$ is taken into account. 

\begin{figure}
\includegraphics[height=4.5cm]{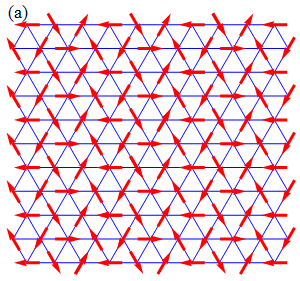}
\includegraphics[height=4.5cm]{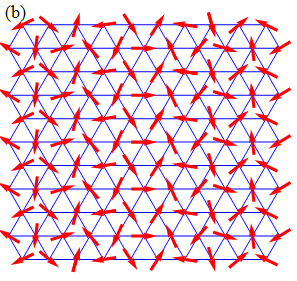} 
\caption{(Color online) (a) Commensurate in-plane spin structure with $\left| {\bf Q}_{comm}\right| = 0.5672\AA^{-1}$. $2\sqrt{3}\times 2\sqrt{3}$ spin structure. (b) Incommensurate in-plane spin structure with $\left| {\bf Q}_{incomm}\right| = 0.522 \AA^{-1}$.}
\label{fig21}
\end{figure}

Using Eq.(\ref{eq24}), the in-plane spin structure with a characteristic {\bf Q} can be obtained on the triangular lattice without knowing the values of $J_{1}$, $J_{2}$, $J_{3}$, and $J^{\prime}$. Figure \ref{fig21}(a) and (b) shows the commensurate in-plane spin structures with $\left| {\bf Q}_{comm}\right| = 0.5672\AA^{-1}$, and incommensurate in-plane spin structure with $\left| {\bf Q}_{incomm}\right| = 0.522\AA^{-1}$. 

\subsection{\label{disB}Nature of reentrant spin glass phase} 
Below $T_{SG}$, the short range in-plane spin order characterized by the incommensurate in-plane wavevector $\tau=\left| {\bf Q}_{incomm}\right|$ coexists with the spin glass phase. The in-plane correlation length slightly increases from $17 \AA$ at $T_{SG}$ to $18 \AA$ ($\approx 5a$) at 0.45 K with decreasing $T$. The wavevector $\tau=\left| {\bf Q}_{incomm}\right|$ decreases from $0.535 \AA^{-1}$ at $T_{SG}$ to $0.522 \AA^{-1}$ at 0.45 K with decreasing $T$, where $\left| {\bf Q}_{comm}\right| = 0.5672 \AA^{-1}$. This means that the degree of the incommensurability of in-plane spin order with the Mn lattice is slightly enhanced well below $T_{SG}$. The in-plane spin correlation length almost remains unchanged below $T_{SG}$ partly because of the highly geometrically frustrated-nature of the 2D AFT. 

The spin glass phase observed in our system is rather different from that of the dilute Ising antiferromagnet Fe$_{0.55}$Mg$_{0.45}$Cl$_{2}$. Wong et al.\cite{ref29} have shown that the spin glass behavior and the long-range antiferromagnetic (AF) order coexist in the low temperature phase. forming the reentrant spin glass phase (RSG). The system undergoes magnetic phase transitions at the N\'{e}el temperature $T_{N}$ = 7.5 K and the reentrant spin glass transition temperature $T_{RSG}$ = 3.0 K. The AF phase with long range spin order exists between $T_{N}$ and $T_{RSG}$ and the RSG phase exits below $T_{RSG}$. The AF order is long range ($>10^{3} \AA$) both between $T_{N}$ and $T_{RSG}$, and below $T_{RSG}$. In other words, it is unaffected by the reentrant spin glass transition. The microscopic picture of the coexistence may consist of spins in an infinite AF network and spins frozen like spin glass. In our system, there are two kinds of spins, spins in the short range spin order and the frustrated spins. The spin direction of the frustrated spins are frozen below $T_{SG}$ because of the spin frustration effect arising from the competing interactions. The short range spin order is not destroyed since this spin order is already established well above $T_{SG}$. 

\section{CONCLUSION} 
The magnetic properties of stage-2 MnCl$_{2}$ GIC are discussed in terms of the experimental results of specific heat as well as ZFC and FC magnetization, AC magnetic susceptibility, magnetic neutron scattering, and ESR. The short range spin order with associated with the incommensurate wave vector $\left| {\bf Q}_{incomm}\right|$ ($= 0.522 \AA^{-1}$ at 0.45K) appears below 5 K ($\approx\left|\Theta_{\perp}\right|$, $\left|\Theta_{\parallel}\right|$), where the magnetic specific heat $C_{mag}$ has a broad peak. The wavevector $\left| {\bf Q}_{incomm}\right|$ is close to the wavevector ($\left| {\bf Q}_{comm}\right|=\sqrt{3}\left| {\bf a}^{*}\right|/6 = 0.5672 \AA^{-1}$) of the commensurate in-plane spin structure with ($2\sqrt{3}\times 2\sqrt{3}$) periodicity. The in-plane spin correlation length slightly increases with decreasing $T$ and reaches only $18 \AA$ at 0.45 K below $T_{SG}$. In the low temperature phase below $T_{SG}$, the spin glass phase coexists with the short-range spin order associated with the incommensurate wave vector $\left|{\bf Q}_{incomm}\right|$. The residual entropy at 0 K is estimated to be 63 \% of the total entropy because of highly frustrated nature of the system. The entropy due to the broad peak around 5 K is 1/3 of the total entropy. The temperature (= 41 K) at which $C_{mag}$ shows a broad peak is the onset temperature of short-range spin order. These features are common to the systems with the same universality class, 2D $XY$ antiferromagnet on the triangular lattice. 

\begin{acknowledgments}
We are grateful to Prof.M. Matsuura, Prof. Y. Kimishima, Dr. D.G. Wiesler, Dr. N. Rosov and Dr. K. Koga for invaluable discussions during their collaborations on the study on the magnetic properties of stage-2 MnCl$_{2}$ GIC since 1980's. The specific heat measurement was done at Tohoku University.  
\end{acknowledgments}

\end{document}